# Hyperspherical-coordinate approach to the spectra and decay widths of hybrid quarkonia

T. Miyamoto[1,2,*] and S. Yasui[1,3,†]

[1]*Department of Physics, Tokyo Institute of Technology, Ookayama 2-12-1, Meguro, Tokyo 152-8550, Japan*
[2]*Graduate School of Engineering Science, Yokohama National University, Tokiwadai 79-1, Yokohama, Kanagawa 240-8501, Japan*
[3]*Research and Education Center for Natural Sciences, Keio University, Hiyoshi 4-1-1, Yokohama, Kanagawa 223-8521, Japan*

In this paper, we consider the possibility that Y(4260), Y(4360), $\psi(4415)$, or X(4660) is the ground state or the first excited state of a hybrid charmonium and examine whether $\Upsilon(10860)$ is the ground state or the first excited state of a hybrid bottomonium. Under a constituent quark model, we carry out numerical calculations to obtain the spectra of both electric gluon and magnetic gluon hybrid quarkonia. We see that the ground state of a magnetic gluon hybrid charmonium and the first excited state of an electric gluon hybrid are comparable in energy to $\psi(4415)$ and that the first excited states of an electric gluon hybrid bottomonium and the ground state of a magnetic gluon hybrid appear a few hundred MeV above the mass of $\Upsilon(10860)$. Also, we evaluate the widths of the decays of a hybrid charmonium into $D^{(*)}\bar{D}^{(*)}$ and those of a hybrid bottomonium into $B^{(*)}\bar{B}^{(*)}$. We show that if the exotic meson candidates are hybrid charmonia, then they are magnetic gluon ones and argue that it seems unlikely that $\Upsilon(10860)$ is the ground state of an electric gluon hybrid bottomonium.

## I. INTRODUCTION

Over the past decades, research on exotic hadrons has been one of the most important areas of research in terms of quantum chromodynamics (QCD), given that the quantum field theory of the strong interaction does not deny the existence of color-singlet particles of this sort. The year 2003 marked the beginning of a new chapter in the history of QCD, because X(3872), the first exotic meson candidate, was observed at Belle [1]. The discovery of X(3872) accelerated the trend of research on exotic hadrons: Since then, various exotic meson candidates have been found, and the possibility of the existence of further exotic hadrons has been raised. This has provided huge motivation to further search for exotic hadrons [2] and led to a better understanding of how the experimentally verified exotic hadron candidates are described.

Among those exotic hadron candidates, Y(4260), which was first observed in 2005, is singled out for special treatment. Using initial state radiation, the *BABAR* Collaboration revealed the existence of the particle [3], and it was subsequently confirmed by the CLEO-c [4] and Belle experiments [5]. Also, it was confirmed that the particle has $J^{PC} = 1^{--}$, because the measured dipion mass distribution agreed with the distribution that was calculated in an S-wave phase space model [6]. There are several reasons why Y(4260) has been regarded as a highly exotic meson: The decay width of the process $Y(4260) \to J/\psi \pi^+ \pi^-$ is significantly large compared to that of other vector charmonia such as $\psi(4040)$; Y(4260)'s decay into $D\bar{D}$ has not been observed; the decay width of the process $Y(4260) \to \psi \pi^+ \pi^-$ was larger than 5.0 MeV, much larger than that of other $1^{--}$ charmonium mesons such as $\psi(3770)$ (80–90 keV) and $\psi(4040)$ ($\leq 100 \pm 30$ keV) [7]; also, Y(4260) decays into an exotic meson Z(3900)$^\pm$ [8].

From a theoretical point of view, there are the following several arguments to explain the anomalous behavior of Y(4260). First, it could be argued that the particle is a $c\bar{c}s\bar{s}$ tetraquark, although the decay channel $Y(4260) \to D_s \bar{D}_s$ has not been observed yet. Second, the possibility that Y(4260) is a $\bar{D}D_1(2420)$ molecule has been raised [9], while there is an argument that the binding energy extracted from the experimental data is stronger than the energy scale that is related to the Yukawa meson exchange interaction. Also, the particle could be interpreted as a hadrocharmonium, that is, a charmonium surrounded by the cloud of light hadrons [10].

[*]miyamoto-tomokazu-gv@ynu.ac.jp, tomokazu.miyamoto10@physics.org
[†]yasuis@keio.jp

There has been another interpretation of the experimental results: The particle is a hybrid charmonium which consists of $c\bar{c}$ and a constituent gluon [11]. As a hybrid meson must conform to its selection rule [12], the ground state of a hybrid charmonium could have either an electric or a magnetic gluon. The decay pattern of Y(4260) suggests that the constituent gluon should be magnetic.

In relation to a charmonium hybrid, let us note that several hybrid mesons have been considered; for the details of light hybrids, see the review paper which was written by Meyer and Swanson [13].

Under the assumption that a charmonium hybrid exists, it is possible that there could exist a bottomonium equivalent of the hybrid meson; several different theoretical studies have explored the possibility of a hybrid bottomonium. For instance, a lattice QCD calculation which employed a nonrelativistic approach showed that the gluonic excitation could exist 1.542 GeV above the "ground" state ($\approx$9.45 GeV) [13,14]. Another lattice calculation, together with the leading Born-Oppenheimer approximation, suggested that the lowest hybrid state could exist 1.49 GeV above the 1S state [15]. Also, the QCD sum rule analyses predicted that there exist several bottomonium hybrids: $J^{PC} = 0^{-+}$ at around $10.6 \pm 0.19$ GeV [16], $1^{--}$ in the mass range of 10.24 to 11.15 GeV [17], and $1^{++}$ at around $11.32 \pm 0.32$ GeV [18]. Furthermore, it has been reported that the spin-dependent structures of hybrid quarkonia were unveiled within the Born-Oppenheimer effective field theory framework [19].

Experimentalists have not yet obtained conclusive evidence that the gluonic degrees of freedom could appear in a bottomonium. However, we can spotlight a few particles which sit above the $B\bar{B}$ threshold: $Y_b(10890)$ is said to be a candidate for a $b\bar{b}g$ [20], and the anomalously large partial width of the decays of $\Upsilon(10860)$ [or $\Upsilon(5S)$] into $\Upsilon(nS)\pi^-\pi^+$ should be looked at more carefully.

Hybrid quarkonia have opened up an opportunity to not only study the gluonic degrees of freedom but also assess how a quark and its antiquark interact in a $q\bar{q}g$. While the $q\bar{q}$ of a usual quarkonium are bound together by the static quark potential, which is not directly measurable in an experiment, a hybrid quarkonium emerges from the excitations of a dynamical gluon. The excitations were studied within the lattice framework [21]; at the same time, it is possible to interpret the gluonic excitations as the string vibration modes under the flux tube regime [22]. The Lüscher term approach can also be used to assess the gluonic excitations [23], which is helpful in placing the constraints on the effective interactions relating to the mixing between hybrids and heavy quarkonium states [24].

In the present paper, we consider the possibility that Y(4260) or one of the other related exotic meson candidates such as Y(4360) is the ground state or the first excited state of a hybrid charmonium and examine whether $\Upsilon(10860)$ is the ground state or the first excited state of a hybrid bottomonium. Under a semirelativistic framework, we calculate the energy spectra of charmonium and bottomonium hybrids and obtain the quark-antiquark effective potentials. Recognizing QCD's nonperturbative nature, we adopt a constituent quark model in which the whole system consists of a quark, its antiquark, and a constituent gluon. Also, we consider the decays of a charmonium hybrid and bottomonium hybrid into $D^{(*)}\bar{D}^{(*)}$ and $B^{(*)}\bar{B}^{(*)}$, respectively, evaluating these decay widths.

Section II consists of several subsections: the auxiliary field approach, hyperspherical formalism, the quark-antiquark effective potentials, and the decays of a hybrid charmonium into $D^{(*)}\bar{D}^{(*)}$ and those of a hybrid bottomonium into $B^{(*)}\bar{B}^{(*)}$. In Sec. III, we present our numerical results. The summary is given in the final section.

## II. FORMALISM

### A. The model for a hybrid meson

In this subsection, we begin by introducing the Hamiltonian of the whole system $q\bar{q}g$, where $q$ and $\bar{q}$ stand for a heavy quark and its antiquark, respectively. According to earlier studies conducted by Kalashnikova and Nefediev [25] (and subsequently, Mathieu [26]), the total Hamiltonian of a hybrid meson is given as the sum of the free Hamiltonian and the potential: $H = H_\text{f} + V$. The free Hamiltonian and the potential term are expressed as follows:

$$H_\text{f} = \sqrt{\mathbf{p}_q^2 + m_q^2} + \sqrt{\mathbf{p}_{\bar{q}}^2 + m_{\bar{q}}^2} + \sqrt{\mathbf{p}_g^2}, \qquad (1)$$

$$V = \sigma r_{qg} + \sigma r_{\bar{q}g} + V_\text{C}, \qquad (2)$$

where $m_q$ and $\mathbf{p}_q$ are the mass and the momentum of a quark, respectively. Likewise, subscripts $\bar{q}$ and $g$ denote an antiquark and a constituent gluon. In this model, the mass $m_g$ of the constituent gluon is strictly zero.

When it comes to the potentials in the Hamiltonian, $\sigma r_{qg}$ and $\sigma r_{\bar{q}g}$ are the linear potentials for the quark confinement, and $V_\text{C}$ is the color-Coulomb potential for this system. In relation to the potentials, the Casimir scaling is considered: Using the SU(3) generators $\{\lambda_i^a\}_{a=1,\ldots,8}$, we have the following relations for a hybrid meson,

$$\frac{1}{4}\langle\lambda_q \cdot \lambda_g\rangle = \frac{1}{4}\langle\lambda_{\bar{q}} \cdot \lambda_g\rangle = -\frac{3}{2}, \qquad \frac{1}{4}\langle\lambda_q \cdot \lambda_{\bar{q}}\rangle = \frac{1}{6}. \qquad (3)$$

Accordingly, we have the expressions for $V_\text{C}$:

$$V_C = -\frac{3\alpha_s}{2r_{qg}} - \frac{3\alpha_s}{2r_{\bar{q}g}} + \frac{\alpha_s}{6r_{q\bar{q}}}. \qquad (4)$$

The Salpeter-type Hamiltonian in (1) contains square roots and so does the corresponding action. In general, this sort of square root has two difficulties: quantization and dealing with a massless particle [27]. Specifically, in the present paper, we need to consider the constituent gluon which is massless. For these reasons, we adopt the auxiliary field method [28] which involves an auxiliary field $\mu$ (or an einbein field) to reparametrize this kind of Hamiltonian:

$$\sqrt{\mathbf{p}^2 + m^2} \Rightarrow \frac{\mathbf{p}^2 + m^2}{2\mu} + \frac{\mu}{2}. \qquad (5)$$

The relation (5) suggests that the physical meaning of the auxiliary field $\mu$ is that it is the effective mass of a particle [29].

Using the auxiliary field method and choosing the center of mass of the three-body system as the coordinate origin, we reparametrize the free Hamiltonian (1) by the auxiliary fields $\mu_q$, $\mu_{\bar{q}}$, and $\mu_g$ of the quark, antiquark, and constituent gluon, respectively,

$$H_f = \frac{\mu_q + \mu_{\bar{q}} + \mu_g}{2} + \frac{\mathbf{p}_q^2 + m_q^2}{2\mu_q} + \frac{\mathbf{p}_{\bar{q}}^2 + m_{\bar{q}}^2}{2\mu_{\bar{q}}} + \frac{\mathbf{p}_g^2}{2\mu_g}, \qquad (6)$$

$$= \frac{\mu_q + \mu_{\bar{q}} + \mu_g}{2} + \frac{m_q^2}{2\mu_q} + \frac{m_{\bar{q}}^2}{2\mu_{\bar{q}}} + \frac{\mathbf{k}_x^2}{2\mu_x} + \frac{\mathbf{k}_y^2}{2\mu_y}, \qquad (7)$$

where $\mu_x$ and $\mu_y$ are defined by

$$\mu_x = \frac{\mu_{\bar{q}}}{2} = \mu_{q\bar{q}}, \qquad \mu_y = \frac{2\mu_{\bar{q}}\mu_g}{2\mu_{\bar{q}} + \mu_g}. \qquad (8)$$

Also, $\mathbf{k}_x$ and $\mathbf{k}_y$ are the momenta which are defined in the three-body Jacobi coordinates; the details of the coordinates are provided in Appendix A.

Strictly speaking, auxiliary fields are dynamical variables, and we need the fields before quantization, but it is difficult to do that; therefore, here we treat the fields ($\mu_q$, $\mu_{\bar{q}}$, and $\mu_g$) as approximate variational parameters [28] in order to determine the ground state of the whole system. We also note that an auxiliary field method leads to, at worst, 7% disagreement between a numerical value and its true value in a numerical calculation [29].

### B. Hyperspherical coordinates

Now we point out that we can take advantage of the auxiliary fields for the linear potentials of (2) in order to treat the reparametrized Hamiltonian (7) as a harmonic oscillator system [30]. However, this sort of analytical treatment does not work properly under the presence of the color-Coulomb force $V_C$, as the interactions cannot be expressed as harmonic oscillators. Admittedly, instead of doing those kinds of analytic calculations, we need to carry out numerical calculations in the present case, adopting a hyperspherical-coordinate approach.

This framework provides a useful tool to study few-body systems such as hadrons and atomic nuclei (see [31,32] for more details); in fact, it was adopted in earlier research on a hybrid charmonium [25]. In this section, we therefore explain the hyperspherical formalism to solve the boundary value problems involving the Hamiltonian.

For convenience, we label each of the constituents of the three-body system as particles 1, 2, and 3, where particles 1 and 2 correspond to a quark and an antiquark, respectively. Particle 3 refers to a constituent gluon in this case. Here we use the Jacobi coordinates for the three-body system; the details of the coordinates are given in Appendix A.

Suppose, in Jacobi coordinates, $X$ is the distance between particles 1 and 2, and $Y$ is the distance between particle 3 and the center of mass of particles 1 and 2. Then, we introduce the hyperspherical coordinates $(\rho, \alpha)$ as follows:

$$x = \rho \cos\alpha = \sqrt{\mu_{12}}X, \qquad y = \rho \sin\alpha = \sqrt{\mu_{12,3}}Y, \qquad (9)$$

where $\rho$ is the hyperradius and $\alpha (\in [0, \frac{\pi}{2}])$ is the hyperangle. For later convenience, we use a shorthand notation $\Omega_5$ to express the set of five angular variables: $\Omega_5 = \{\alpha, \theta_x, \phi_x, \theta_y, \phi_y\}$, where $\theta_x$ and $\phi_x$ are the polar and azimuthal angles of the $\mathbf{X}$ vector, respectively. Similarly, $\theta_y$ and $\phi_y$ are the polar and azimuthal angles of the $\mathbf{Y}$ vector, respectively.

The hyperspherical harmonics (HH) is equipped with its related quantum numbers: $K$ is called the hyperangular momentum, $L_x$ is the orbital angular momentum between particles 1 and 2, and $L_y$ is that between particle 3 and the center of mass of particles 1 and 2. Also, an integer $n$ is defined by

$$K = L_x + L_y + 2n. \qquad (10)$$

With these quantum numbers, we introduce the HH for the three-body system,

$$\mathcal{Y}_{KLM_L}^{L_xL_y}(\Omega_5) \stackrel{\text{def}}{=} \mathcal{N}(K, L_x, L_y) A_n^{L_xL_y}(\alpha) \\ \times [Y_{L_x}(\Omega_x) \otimes Y_{L_y}(\Omega_y)]_{L_{\text{tot}}M_{L_{\text{tot}}}}, \qquad (11)$$

where we define the hyperangular part $A_n^{L_xL_y}(\alpha)$ by using the Jacobi polynomials $\{P_n^{\alpha,\beta}\}$ as follows:

$$A_n^{L_xL_y}(\alpha) = (\cos\alpha)^{L_x}(\sin\alpha)^{L_y} P_n^{L_y+1/2, L_x+1/2}(\cos 2\alpha). \qquad (12)$$

Here, $\mathcal{N}(K, L_x, L_y)$ which appears in (11) is a normalization constant given by

$$\mathcal{N}(K,L_x,L_y) = \sqrt{2(K+2)}\sqrt{\frac{\Gamma(n+1)\Gamma(L_x+L_y+n+2)}{\Gamma(L_x+n+\frac{3}{2})\Gamma(L_y+n+\frac{3}{2})}}. \quad (13)$$

In the present method, the total wave function is expanded with regard to the HH,

$$\Psi_{JM_J}(\mathbf{x},\mathbf{y}) = \frac{1}{\rho^{5/2}}\sum_{K\gamma}\chi_{K\gamma}(\rho)[\mathcal{Y}^{L_xL_y}_{KL_{\mathrm{tot}}}(\Omega_5)\otimes|S_{\mathrm{tot}}\rangle]_{JM_J}$$

$$= \frac{1}{\rho^{5/2}}\sum_{K\gamma}\chi_{K\gamma}(\rho)\Upsilon_{K\gamma JM_J}, \quad (14)$$

where $\{\chi_{K\gamma}(\rho)\}_{K,\gamma}$ are the hyperradial functions of the whole wave function, $|S_{\mathrm{tot}}M_{S_{\mathrm{tot}}}\rangle$ is the spin part, and $\gamma$ is a set of quantum numbers: $\gamma = \{L_{\mathrm{tot}}, S_{\mathrm{tot}}, L_x, L_y\}$. We also note that $\Upsilon_{K\gamma JM_J}(\Omega_5)$ is given by $\Upsilon_{K\gamma JM_J}(\Omega_5) = [\mathcal{Y}^{L_xL_y}_{KL_{\mathrm{tot}}}\otimes|S_{\mathrm{tot}}\rangle]_{JM_J}$.

In actual calculations, we need to numerically obtain the hyperradial functions. To do this, we substitute (14) into the Schrödinger equation, which yields the following set of inhomogeneous differential equations (hyperradial equations):

$$\frac{-1}{2}\left(\frac{\partial^2}{\partial\rho^2} - \frac{(K+3/2)(K+5/2)}{\rho^2}\right)\chi_{K\gamma}$$
$$+ \sum_{K\gamma}V_{K\gamma K',\gamma'}\chi_{K'\gamma'} = E\chi_{K\gamma}, \quad (15)$$

where the matrix element of a potential is defined by

$$V_{K\gamma,K'\gamma'} \overset{\mathrm{def}}{=} \langle\Upsilon_{K\gamma}(\Omega_5)|V_{12}+V_{23}+V_{31}|\Upsilon_{K'\gamma'}(\Omega_5)\rangle. \quad (16)$$

Thus, we see that the three-body problem is reduced to dealing with a one-dimensional problem, but we are still confronted with the inhomogeneous equations. Here we address the problem by expanding each of the hyperradial functions with regard to a generalized Laguerre-based basis function in one-dimensional space:

$$\chi_{K\gamma} = \sum_{j=0}^{B-1}a_j^{(m)}u_{\alpha,j}(\rho), \quad (17)$$

where $B$ is the number of the basis functions. The generalized Laguerre-based basis functions $\{u_{\alpha,j}\}$ are defined by

$$u_{\alpha,j}(z) = \sqrt{\frac{\Gamma(j+1)}{\Gamma(\alpha+j+1)}}e^{-\frac{z}{2}}z^{\frac{\alpha}{2}}L_j^{(\alpha)}(z). \quad (18)$$

By virtue of this, our problem is reduced to solving the boundary value problems which involve $a_j^{(m)}$. We note that here $m$ is a set of quantum numbers: $m = \{K,\gamma\}$. In the present case, we set $\alpha = 5$.

The whole wave function $\Psi$ is normalized as

$$\int|\Psi_{JM_J}|^2 d\mathbf{x}d\mathbf{y} = \sum_{K\gamma}\int_0^\infty|\chi_{K\gamma}|^2 d\rho = 1. \quad (19)$$

### C. Effective potential of a color-octet quark-antiquark pair

As mentioned earlier, standard quarkonium states emerge from the static quark potential, while the excitations of dynamic gluons are responsible for hybrid meson states. From a constituent quark model perspective, the gluonic excitations are described by the effective potential with which the constituent gluon of a $q\bar{q}g$ involves, keeping the quark-antiquark subsystem color octet. In other words, the information about the gluonic degrees of freedom can be accessed through studying the quark-antiquark effective potential.

In this subsection, therefore, we calculate the quark-antiquark effective potentials in a $q\bar{q}g$. The effective potential is derived from integrating over the coordinates $\mathbf{Y}$ relating to the constituent gluon of a hybrid meson:

$$\hat{V}(\mathbf{X}) = \langle g|V(\mathbf{X},\mathbf{Y})|g\rangle, \quad (20)$$

where coordinates $\mathbf{X}$ represent the $q\bar{q}$ relative position.

Our approach to derive the effective potential is to take advantage of the Lagrange mesh method (LMM). Details of the LMM are given in Appendix C. Suppose that the eigenenergy $E$ and the whole wave function of a physical system are given. Then, the procedure for calculating the potentials consists of two parts: We extract a two-body radial wave function $R(X)$ from the whole wave function, then we obtain the effective potential relating the subsystem by solving the inverse problem.

The two-body radial wave function $R(X)$ of a color-octet subsystem $q\bar{q}$ is obtained by solving the following equation:

$$R(X) = \int|\Psi(\mathbf{X},\mathbf{Y})|^2 d\mathbf{Y}d\Omega_X, \quad (21)$$

where $\Omega_X$ is a shorthand notation for the angular variables of the $\mathbf{X}$ vector. Then, after normalizing $R(X)$, we obtain a dimensionless wave function by $\phi_X(X) = XR(X)$. By using the dimensionless two-body radial wave function, we solve the inverse problems in terms of the Hamiltonian of the subsystem in order to obtain the effective potential,

$$\hat{V}(x_i) = E - \sum_{j=1}^N T_{ij}\frac{\sqrt{\lambda_j}\phi(x_j)}{\sqrt{\lambda_i}\phi(x_i)}, \quad (22)$$

where $\{T_{ij}\}_{i,j}$ are the matrix elements of the Hamiltonian, and $\{x_i\}_i$ and $\{\lambda_i\}_i$ are defined in Appendix C.

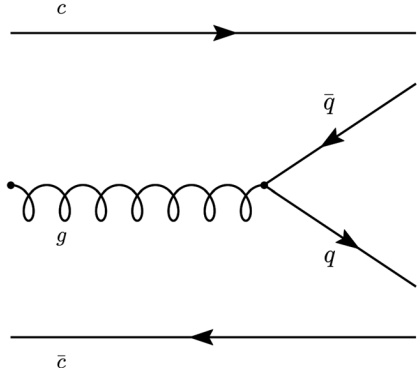

FIG. 1. $D\bar{D}$ decay channel of a $c\bar{c}g$.

### D. Decays of hybrid mesons

Now we consider the decays of a hybrid charmonium into $D^{(*)}\bar{D}^{(*)}$. Under the selection rules, a magnetic gluon $c\bar{c}g$ does not get involved in these processes [11], and therefore, we consider only electric gluon hybrids.

In the present case, we consider the lowest-order decay, as is shown in Fig. 1. Its formalism was already presented in earlier research [12,33], where the wave functions of the initial and final particles were described by Gaussian-type functions. The final states are treated in a nonrelativistic way, given the fact that the mass of a D meson is at lightest about 1.86 GeV. Also, we note that this method could be more suitable for calculating the widths of the decays of a $b\bar{b}g$ into $B^{(*)}\bar{B}^{(*)}$ because the final mesons are much heavier so that the nonrelativistic approximation is more reasonable.

The initial and final wave functions of a $c\bar{c}g$ influence the decay amplitude $f_{A\to BC}$ (whose details are given in [33]). In the present work, obtaining the wave functions of an initial state, we use the momentum representation to calculate the decay widths (for details of the representation, see Appendix D). Also, we take into account the effects of the correlation between the charm-anticharm pair and the constituent gluon of a $c\bar{c}g$.

Suppose a $c\bar{c}g$ is particle A, and D and $\bar{D}$ mesons are particles B and C, respectively. Here we denote their masses by $m_A$, $m_B$, and $m_C$. Then, the decay width $\Gamma_{A\to BC}$ of the process is expressed as

$$\Gamma_{A\to BC} = 4\alpha_s |f_{A\to BC}|^2 \frac{P_B E_B E_C}{m_A}, \quad (23)$$

where $E_B$ and $E_C$ are the energies of particles B and C, respectively: $E_B = \sqrt{P_B^2 + m_B^2}$ and $E_C = \sqrt{P_B^2 + m_C^2}$. We also note that here $P_B$ is given by

$$P_B = \frac{1}{2m_A} \cdot \sqrt{\{m_A^2 - (m_B + m_C)^2\}\{m_A^2 - (m_B - m_C)^2\}}. \quad (24)$$

TABLE I. Possible low-lying states for a hybrid meson. In the gluon column, E and M mean the electric gluon and magnetic gluon, respectively.

| Gluon | $L_{q\bar{q}}$ | $L_g$ | $L_{\text{tot}}$ | $S_{q\bar{q}}$ | $S_g$ | $S_{\text{tot}}$ | $J_{q\bar{q}}$ | $J_g$ | $J^{PC}$ |
|---|---|---|---|---|---|---|---|---|---|
| E | 1 | 0 | 1 | 1 | 1 | 0, 1, 2 | $(0,1,2)^{++}$ | $1^-$ | $1^{--}$ |
| M | 0 | 1 | 1 | 0 | 1 | 1 | $0^{-+}$ | $1^+$ | $1^{--}$ |

### III. RESULTS

In this section, we present the results of our numerical calculations with respect to hybrid quarkonia. We begin by distinguishing the magnetic gluon case from the electric case using the parity $P$ and parity $C$ of a hybrid meson: They are given by $P = (-1)^{L_{q\bar{q}} + L_g}$, $C = (-1)^{L_{q\bar{q}} + S_{q\bar{q}} + 1}$; also, we have the relation of $L_g = J_g$ for a magnetic gluon and $L_g = J_g \pm 1$ for an electric gluon [12]. Taking these things into account, we show in Table I the low-lying states which are allowed to exist for a hybrid meson $q\bar{q}g$.

For the magnetic gluon $q\bar{q}g$, we restrict ourselves to the case where $S_{q\bar{q}} = 0$ and $J^{PC} = 1^{--}$, mentioning that $S_{q\bar{q}} = 1$ leads to $J^{PC} = (0,1,2)^{-+}$ [26]. The specific values of the other quantum numbers such as $L_{q\bar{q}}$ are shown in Table I. Strictly speaking, we need to consider higher values for the quantum numbers, but here we stick to the approximations. Similarly, we consider only $S_{q\bar{q}} = 1$ for the electric gluon, pointing out that, in this case, $J^{PC} = 1^{--}$ does not involve the spin singlet state of the quark-antiquark pair.

#### A. Charmonium hybrid

We now focus on the spectrum of a charmonium hybrid $c\bar{c}g$. As to the parameters relating to the potentials in the Hamiltonian, we set the string tension $\sigma$ and color-Coulomb parameter $\alpha_s$ 0.16 GeV$^2$ and 0.55, respectively. Here we used 1.48 GeV for the mass $m_c$ of the charm quark [25]. This parameter setting reproduced the energy spectrum of the charmonium to the extent that the difference between our calculation and the experimental data was less than 1%.

Then, taking advantage of the auxiliary field method, we obtain the effective masses ($\mu_c = \mu_{\bar{c}}$ and $\mu_g$) of a charm, anticharm, and constituent gluon. The results are shown in Fig. 2, where the calculated ground state masses are plotted as functions of $\mu_c$ and $\mu_g$. As to the values of $\gamma = \{L_{\text{tot}}, S_{\text{tot}}, L_{c\bar{c}}, L_g\}$, we used Table I to determine them according to the type of constituent gluon (i.e., electric or magnetic). Here we retained the maximum of the hyperangular momentum $K$ up to 9, which allowed us to see a reasonable convergence (see Fig. 3).

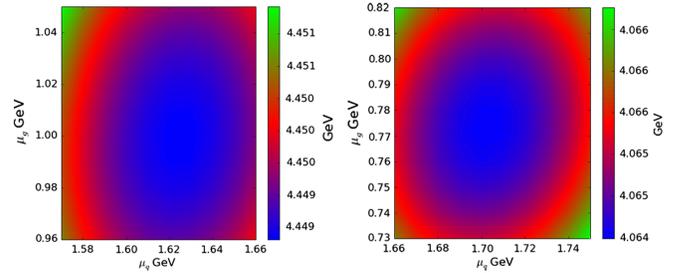

FIG. 2. The calculated ground state masses (GeV) of a $c\bar{c}g$ as functions of the auxiliary fields $\mu_c$ and $\mu_g$. The left graph is for the magnetic gluon case and the right is for the electric gluon case.

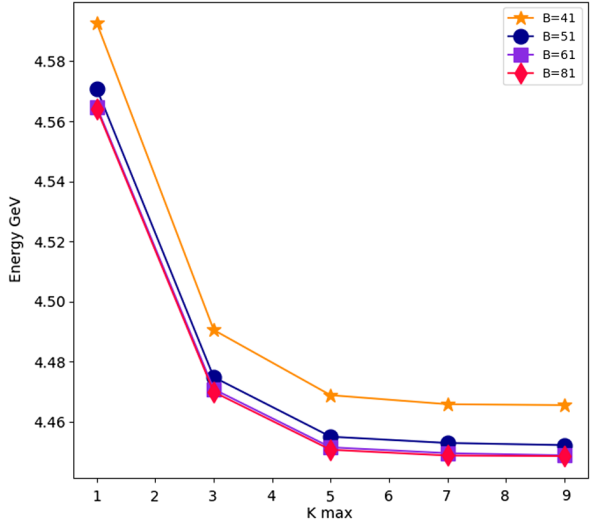

FIG. 3. The ground state energies of $c\bar{c}g$ as a function of the maximum of $K$. $B$ is the number of the generalized Laguerre-based basis functions.

Then, the resulting basic parameters of the spectrum of a $c\bar{c}g$ are shown in Table II.

In earlier research, a single Gaussian was used as their basis function [25], while we used multicomponent generalized Laguerre-based functions to express the hyperradial functions. For this reason, the ground state energy that ($\approx 4.448$ GeV) we obtained was lower than theirs (4.573 GeV) [26]. In relation to this, we mention that if we use a single-component basis function, then the ground state energy is consistent with their result.

When it comes to the Gaussian correlated approach, the ground state energy for the magnetic case was 4.445 GeV [26]. Figure 3 shows that our results are consistent with this.

Next we present the mass spectra of a $c\bar{c}g$ in the right part of Fig. 4. From the graph, it is apparent that for the electric gluon case we have lower masses as a whole, compared to the magnetic gluon case. We can see that the ground state of an electric gluon $c\bar{c}g$ appears about

TABLE II. The parameters relating to the spectra of a $c\bar{c}g$. E and M in the gluon column of the table mean the electric gluon case and magnetic gluon case, respectively.

| $\sigma$ (GeV$^2$) | $\alpha_s$ | $m_c$ (GeV) | $J^{PC}$ | Gluon | $\mu_c$ (GeV) | $\mu_g$ (GeV) |
|---|---|---|---|---|---|---|
| 0.16 | 0.55 | 1.48 | $1^{--}$ | M | 1.62 | 1.00 |
| 0.16 | 0.55 | 1.48 | $1^{--}$ | E | 1.70 | 0.77 |

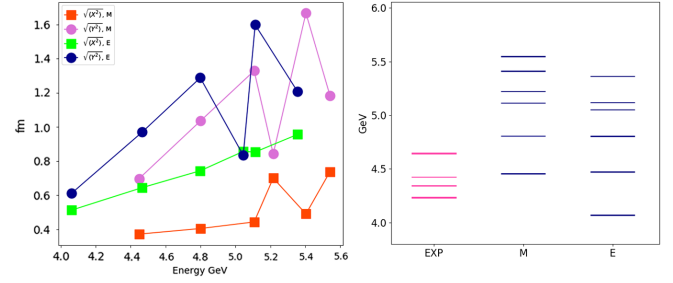

FIG. 4. In the left graph, the rms radii $\sqrt{\langle X^2 \rangle}$ and $\sqrt{\langle Y^2 \rangle}$ for a $c\bar{c}g$ are plotted against its eigenenergies. In the right graph, the relevant experimental data [Y(4260), Y(4360), $\psi$(4415), and X(4660)] and the calculated energy spectra are shown [34], where E stands for the electric case and M stands for the magnetic case.

200 MeV below the mass of Y(4260), while the first excited state of an electric gluon hybrid and the ground state of a magnetic gluon hybrid are comparable in energy to $\psi$(4415).

Then we consider the relation between the mass and rms radius in terms of a charmonium hybrid for both the magnetic gluon and electric gluon. In the left part of Fig. 4, the rms radii $\sqrt{\langle X^2 \rangle}$ and $\sqrt{\langle Y^2 \rangle}$ for a $c\bar{c}g$ are plotted against the masses of the hybrid, where $X$ is the distance between $c$ and $\bar{c}$, and $Y$ is the distance between the center of mass of $c\bar{c}$ and a constituent gluon. We can see on this graph that $\sqrt{\langle X^2 \rangle}$ for the electric case is, as a whole, larger than that for the magnetic case; the same thing holds true for $\sqrt{\langle Y^2 \rangle}$. This result is consistent with our findings on the spectra.

Also, the probability densities of a $c\bar{c}g$ for its ground states are shown in Fig. 5, where the density distributions are scaled so that their peaks and bottoms are set to unity and zero, respectively. On these graphs, we see the broad configuration in the Y direction for the magnetic case, while the density distribution has a rather long tail in the X direction for the electric case. These spatial configurations are consistent with the fact that a constituent gluon sits in the P state, and the orbital angular momentum between $c$ and $\bar{c}$ is zero for the magnetic case, and that a constituent gluon sits in the S state, and the orbital angular momentum between $c$ and $\bar{c}$ is 1 for the electric case.

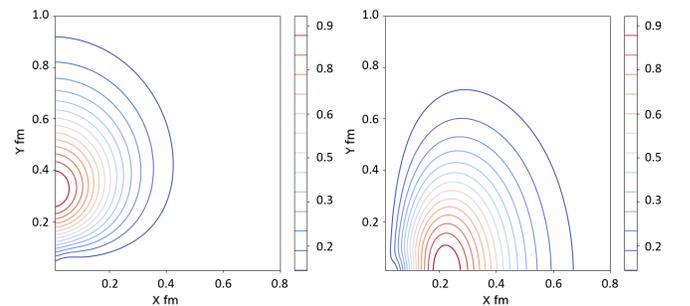

FIG. 5. The scaled probability densities of the ground states of a $c\bar{c}g$ for the magnetic case (left) and the electric case (right). The density distributions are scaled so that their peaks are unity and their bottoms are zero.

### B. Bottomonium hybrid

We then move on to consider a heavier hybrid meson $b\bar{b}g$. Before calculating the energy spectra of a hybrid bottomonium, we need to obtain relevant parameters: the mass $m_b$ of a bottom quark and the color-Coulomb parameter $\alpha_s$. These parameters are fitted to the experimental data in terms of the energy spectrum of a bottomonium. The details of the fitting procedures are explained in Appendix B.

Using the parameters $m_b$ and $\alpha_s$ that were obtained, we calculate the ground state energies of a $b\bar{b}g$ through the auxiliary field method.

The basic inputs and outputs are summarized in Table III. From the table, it is apparent that the calculated total energies of a bottomonium hybrid are about 10.6 and 11.2 GeV for the electric gluon case and magnetic case, respectively. These results are consistent with those of the QCD sum rule analyses [17].

Also, the effective mass of the constituent quark of a $b\bar{b}g$ is in the range of 4.9 to 5.1 GeV, much heavier than that ($\sim 1.6$ GeV) of a $c\bar{c}g$. In contrast, the constituent gluon of a hybrid bottomonium is comparable to that of a hybrid charmonium in terms of effective mass: $\mu_g$ is in the range of 0.9 to 1.3 GeV, and $\mu_g$ is in the range of 0.7 to 1.0 GeV for a $c\bar{c}g$.

Using these expectation values of the auxiliary fields $\mu_b$ and $\mu_g$, we calculated the energy spectra of a bottomonium hybrid. The calculated energy spectra of a $b\bar{b}g$ are shown in the right part of Fig. 6, where we can see that our theoretical values are deviated from the mass of $\Upsilon(10860)$ for both the magnetic and electric cases. In the left part of Fig. 6, we plot the rms radii of X and Y against the eigenenergies. From the graph, we can see that basically these rms radii for the electric case are larger than those for the magnetic case, which is consistent with the fact that the calculated eigenenergies of a bottomonium hybrid for the magnetic case are larger than those for the electric case.

The probability densities of a bottomonium hybrid for its ground states are shown in Fig. 7, where we can see the similar spatial configurations to those of a $c\bar{c}g$. This reflects the fact that $L_g = 1$, and the orbital angular momentum between $b$ and $\bar{b}$ is zero for the magnetic case, and that $L_g = 0$ and the orbital angular momentum between $b$ and $\bar{b}$ is 1 for the electric case.

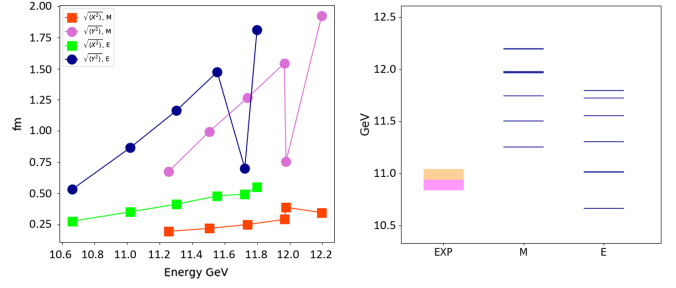

FIG. 6. In the left graph, the rms radii of X and Y are plotted against the eigenenergies for both the magnetic and electric gluon cases. In the right graph, the theoretical energy spectra of a $b\bar{b}g$ are shown, where M and E are the magnetic gluon and electric gluon, respectively. The experimental data on the spectra [of $\Upsilon(10860)$ and $\Upsilon(11020)$] are shown together [34].

Also, we show the quark-antiquark effective potentials for the first three states of a bottomonium hybrid in Fig. 8, where both the magnetic and electric cases are considered, and the lattice calculations ($\Sigma^{+,ex}, \Pi_g$) are shown together [21]. (For the details of this greek letter notation and the related sub- or superscripts, see [21].) Here we show the effective potentials only for set A as they are not qualitatively different from those for set B. The calculated potentials are plotted as a function of a dimensionless quantity $X/R_0$. Here, $R_0$ is the hadronic scale parameter which is defined by $r^2 \frac{dV_{q\bar{q}}(\mathbf{r})}{dr}|_{r=R_0} = 1.65$, where $V_{q\bar{q}}$ is the static quark potential [35]. The value of the parameter is about 0.5 fm [35]; in the present case, we set the parameter at 0.4933 fm. We also note that the vertical axis in those graphs is made dimensionless: $R_0(V_{\text{eff}}(X) - V_{q\bar{q}}(2R_0))$.

We estimate that, for the magnetic cases, the our calculations' numerically reliable regions are $X/R_0 \in [0.4, 0.8]$ for the first two states and $[0.5, 0.8]$ for the second excited state. For the electric cases, we estimate the reliable regions are $X/R_0 \in [0.4, 1.4]$. The calculated potentials are smaller than those of the lattice calculations, especially for the electric cases.

TABLE III. The ground state energies of a bottomonium hybrid and its relevant parameters. In the gluon column, M stands for magnetic and E stands for electric.

| | $\sigma$ (GeV$^2$) | $\alpha_s$ | $m_b$ (GeV) | $J^{PC}$ | Gluon | $\mu_b$ (GeV) | $\mu_g$ (GeV) | GS energy (GeV) |
|---|---|---|---|---|---|---|---|---|
| Set A | 0.16 | 0.55 | 5.00 | $1^{--}$ | M | 5.06 | 1.27 | 11.2 |
| | 0.16 | 0.55 | 5.00 | $1^{--}$ | E | 5.17 | 1.10 | 10.6 |
| Set B | 0.16 | 0.45 | 4.85 | $1^{--}$ | M | 4.90 | 1.14 | 11.1 |
| | 0.16 | 0.45 | 4.85 | $1^{--}$ | E | 4.98 | 0.94 | 10.6 |

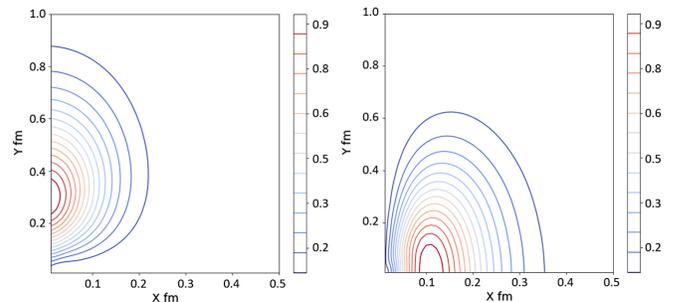

FIG. 7. The probability densities of a $b\bar{b}g$ for its ground states. The left is for the magnetic gluon and the right is for the electric gluon. The density distributions are scaled so that their tops are unity and their bottoms are zero.

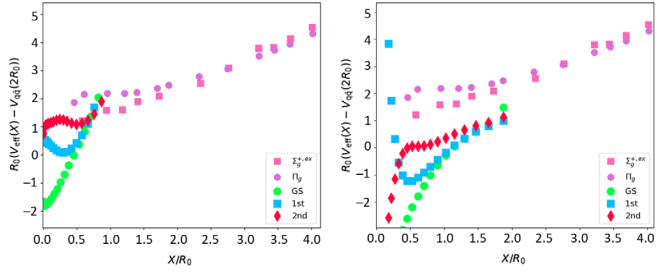

FIG. 8. The calculated quark-antiquark effective potentials (set A) for the first three states of a $b\bar{b}g$. The left graph is for the magnetic gluon case and the right is for the electric case. The lattice calculations ($\Sigma^{+,ex}, \Pi_g$) are shown together [21].

### C. Partial decay widths

Now we calculate the decay widths for the processes of a $c\bar{c}g$ (whose constituent gluon is electric) into $D^{(*)}\bar{D}^{(*)}$. Before doing this, we set the relevant parameters, which are given in Table IV, where $\omega$ is the averaged energy difference between the ground state and the first excited state. $\mu_D$ and $\mu_B$ are the reduced masses which are defined by

$$\mu_D = \frac{m_c m_{\bar{q}}}{m_c + m_{\bar{q}}}, \qquad \mu_B = \frac{m_b m_{\bar{q}}}{m_b + m_{\bar{q}}}. \qquad (25)$$

Also, introducing a Gaussian-related parameter $R$, we have the following relation [33]:

$$R^2 = \frac{1}{\omega\mu}. \qquad (26)$$

Then, we summarize these parameters in Table IV.

In this case, we set $m_q = m_{\bar{q}} = 0.35$ GeV, where $q = u$, $d$, and $m_s = 0.5$ GeV [33,36]. We show the results in Table V, where we consider the decays of a $c\bar{c}g$ into $D\bar{D}$, $D_s^+ D_s^-$, $D^*\bar{D} + D\bar{D}^*$, and $D^*\bar{D}^*$. We note that $D\bar{D}$ means the sum of $D^0\bar{D}^0$ and $D^+D^-$; similarly, $D^*\bar{D} + D\bar{D}^*$ consists of $D^{*0}\bar{D}^0 + D^0\bar{D}^{*0}$ and $D^{*+}D^- + D^+D^{*-}$ and also $D^*\bar{D}^* = D^{*0}\bar{D}^{*0} + D^{*+}D^{*-}$. Furthermore, let us denote all of these channels by $D^{(*)}\bar{D}^{(*)}$.

From the table, we see that the partial widths of the decays from the ground state to these states are in the range of 0.01 to 0.65 GeV, and that the summations of those widths are about 0.60, 0.77, and 1.15 GeV for $J_{q\bar{q}} = 0$, 1, and 2, respectively. The calculated total widths are much larger than the experimental data (~0.05 GeV) [34]. This result suggests that if the exotic meson candidate is a charmonium hybrid, it is unlikely that the $c\bar{c}g$'s constituent gluon is electric. Therefore, Y(4260) contains a magnetic gluon if it is a hybrid charmonium, which is in line with earlier research [11].

As to the decays from the first excited state of a $c\bar{c}g$ to $D^{(*)}\bar{D}^{(*)}$, the summations of the partial widths are about 0.18, 0.22, and 0.29 GeV for $J_{q\bar{q}} = 0$, 1, and 2, respectively. Although these values are smaller than those for the decays from the ground state of a $c\bar{c}g$, they are still much larger than the observed total widths (~0.1, ~0.06, and ~0.07 GeV) of Y(4360), $\psi$(4415), and X(4660), respectively. This result supports the view that these exotic meson candidates are not electric gluon hybrid mesons.

TABLE IV. The relevant parameters for the decay processes of hybrid mesons into nonexotic mesons. Here, $m_c$ and $m_b$ are the masses of a charm quark and bottom quark, respectively. The values in parentheses stand for the values for the cases where a strange quark is involved.

|  |  | $m_c$ or $m_b$ (GeV) | $\omega$ (GeV) | $\mu_D$ or $\mu_B$ (GeV) | $R$ (GeV$^{-1}$) |
|---|---|---|---|---|---|
| $D$ meson |  | 1.48 | 0.468 | 0.283 (0.373) | 2.74 (2.39) |
| $B$ meson | set A | 5.00 | 0.446 | 0.327 (0.454) | 2.61 (2.22) |
|  | set B | 4.85 | 0.446 | 0.326 (0.453) | 2.62 (2.22) |

TABLE V. The calculated partial widths (MeV) of the decays of a charmonium hybrid into $D^{(*)}\bar{D}^{(*)}$. Here the hybrid meson's constituent gluon is electric. We note that $D\bar{D} = D^0\bar{D}^0 + D^+D^-$, $D\bar{D}^* + D^*\bar{D}$ means $D^{*0}\bar{D}^0 + \bar{D}^{*0}D^0$ and $D^{*+}D^- + D^+D^{*-}$. Also, $D^*\bar{D}^* = D^{*0}\bar{D}^{*0} + D^{*+}D^{*-}$.

|  |  | $D\bar{D}$ | $D^*\bar{D} + D\bar{D}^*$ | $D^*\bar{D}^*$ | $D_s^+ D_s^-$ | Total |
|---|---|---|---|---|---|---|
| $\Gamma_{GS \to D^{(*)}\bar{D}^{(*)}}$ | $J_{q\bar{q}}$ |  |  |  |  |  |
|  | 0 | 123.36 | 344.41 | 127.25 | 13.49 | 608.5 |
|  | 1 | 370.08 | 258.30 | 109.07 | 40.48 | 777.9 |
|  | 2 | 616.80 | 430.51 | 36.35 | 67.46 | 1151.1 |
| $\Gamma_{1st \to D^{(*)}\bar{D}^{(*)}}$ | $J_{q\bar{q}}$ |  |  |  |  |  |
|  | 0 | 28.87 | 81.19 | 70.53 | 6.64 | 187.2 |
|  | 1 | 86.61 | 60.89 | 60.45 | 19.94 | 227.9 |
|  | 2 | 144.36 | 101.48 | 20.15 | 33.23 | 299.2 |

TABLE VI. The decay widths (MeV) of the processes of $b\bar{b}g \to B^{(*)}\bar{B}^{(*)}$ are shown, where the experimental data of the decay of $\Upsilon(10860)$ into $B^{(*)}\bar{B}^{(*)}$ are shown together [34]. Here the hybrid meson's constituent gluon is electric. The relevant notations are as follows: $B\bar{B}$ stands for $B^0\bar{B}^0 + B^+B^-$, $B^*\bar{B} + B\bar{B}^*$ means $B^{*0}\bar{B}^0 + B^0\bar{B}^{*0}$ and $B^{*+}B^- + B^+B^{*-}$, and $B^*\bar{B}^*$ consists of $B^{*0}\bar{B}^{*0}$ and $B^{*-}B^{*+}$.

| | | | $B\bar{B}$ | $B^*\bar{B}+B\bar{B}^*$ | $B^*\bar{B}^*$ | $B_s\bar{B}_s$ | $B_s\bar{B}_s^*+B_s^*\bar{B}_s$ | $B_s^*\bar{B}_s^*$ | Total |
|---|---|---|---|---|---|---|---|---|---|
| Set A | $\Gamma_{GS\to B^{(*)}\bar{B}^{(*)}}$ | $J_{q\bar{q}}$ | | | | | | | |
| | | 0 | 77.01 | 158.36 | 34.02 | (…) | (…) | (…) | 269.4 |
| | | 1 | 231.05 | 118.77 | 29.16 | (…) | (…) | (…) | 379.0 |
| | | 2 | 385.09 | 197.95 | 9.72 | (…) | (…) | (…) | 592.8 |
| | $\Gamma_{1st\to B^{(*)}\bar{B}^{(*)}}$ | | | | | | | | |
| | | 0 | 5.76 | 12.05 | 6.84 | 0.91 | 0.72 | 0.082 | 26.4 |
| | | 1 | 17.29 | 9.04 | 5.86 | 2.75 | 0.54 | 0.070 | 35.6 |
| | | 2 | 28.82 | 15.06 | 1.95 | 4.59 | 0.91 | 0.023 | 51.4 |
| Set B | $\Gamma_{GS\to B^{(*)}\bar{B}^{(*)}}$ | | | | | | | | |
| | | 0 | 16.31 | (…) | (…) | (…) | (…) | (…) | 16.3 |
| | | 1 | 48.95 | (…) | (…) | (…) | (…) | (…) | 49.0 |
| | | 2 | 81.58 | (…) | (…) | (…) | (…) | (…) | 81.6 |
| | $\Gamma_{1st\to B^{(*)}\bar{B}^{(*)}}$ | | | | | | | | |
| | | 0 | 14.26 | 37.22 | 34.47 | 1.56 | 1.77 | 0.17 | 89.5 |
| | | 1 | 42.80 | 27.92 | 29.55 | 4.68 | 1.32 | 0.14 | 106.4 |
| | | 2 | 71.34 | 46.53 | 9.85 | 7.81 | 2.21 | 0.04 | 137.8 |
| $\Gamma_{exp}(\Upsilon(10860))$ | | | $2.80\pm0.51$ | $6.98\pm0.81$ | $19.43\pm1.73$ | $(2.5\pm0.2)\times10^{-3}$ | $0.69\pm0.16$ | $8.97\pm1.37$ | |

Next, we consider the decays of a $b\bar{b}g$ into $B^0\bar{B}^0$, $B^+B^-$, $B^*\bar{B}$, $B^*\bar{B}^*$, $B_s\bar{B}_s$, $B_s\bar{B}_s^*$, and $B_s^*\bar{B}_s^*$. The calculated decay widths are presented in Table VI, where we use the following notations: $B\bar{B}$ stands for $B^0\bar{B}^0 + B^+B^-$, $B^*\bar{B}+B\bar{B}^*$ means $B^{*0}\bar{B}^0 + B^0\bar{B}^{*0}$ and $B^{*+}B^- + B^+B^{*-}$, and $B^*\bar{B}^*$ consists of $B^{*0}\bar{B}^{*0}$ and $B^{*-}B^{*+}$. Also, the experimental data are shown together in the table [34].

As to the decay of a bottomonium hybrid which is in the ground state into $B^{(*)}\bar{B}^{(*)}$, it is apparent from the table that, for set A, the sum of the partial widths is much larger than $\Upsilon(10860)$'s total decay width (~51 MeV). Although for set B's ground state, the sum of the partial widths is comparable to $\Upsilon(10860)$'s total width, the set of parameters does not allow a $b\bar{b}g$ to decay into the states that involve $B^*$ or $B_s$. For the decay into $B^0\bar{B}^0$ or $B^+B^-$, there are significant differences between the calculated values and the experimental data.

When it comes to the first excited state into $B^{(*)}\bar{B}^{(*)}$, for both sets A and B, the sums of the partial widths are comparable to $\Upsilon(10860)$'s total width except that for $J_{q\bar{q}} = 1$ and 2, the set B yields the total widths which exceed the measured value by more than 100%. However, our partial widths are in the order of $10^3$ times greater than the experimental data in terms of the decays into $B_s\bar{B}_s$; the calculated widths are, on average, roughly $10^{-2}$ of the measured width for the $B_s^*\bar{B}_s^*$ channel. On top of that, for $J_{q\bar{q}} = 1$ and 2, the calculated $B\bar{B}$ partial widths are greater than the calculated $B^*\bar{B}^*$ partial widths, while we see the opposite in the experimental data.

Our results suggest that it seems unlikely that $\Upsilon(10860)$ is the ground state of an electric gluon hybrid bottomonium and that the argument that $\Upsilon(10860)$ is the first excited state of an electric gluon $b\bar{b}g$ is weak. Having said that, we do not rule out the possibility that the first excited state of a bottomonium hybrid will be discovered as a state which is different from $\Upsilon(10860)$ since the sum of the partial widths (for set A) is around 35 MeV.

The parameter setting with regard to $\alpha_s$ is a matter of debate because the running coupling constant $\alpha_s$ used in (23) should be smaller than that used in (4). If we take the difference in energy scale into consideration, the deviations from the experimental data become smaller with respect to the channels of $B\bar{B}$, $B^*\bar{B}+B\bar{B}^*$, $B_s\bar{B}_s$, and $B_s\bar{B}_s^*+B_s^*\bar{B}_s$.

## IV. SUMMARY

In this work, we considered the possibility that Y(4260), Y(4360), $\psi$(4415), or X(4660) is the ground state or the first excited state of a hybrid charmonium and also examined whether $\Upsilon(10860)$ is the ground state or the first excited state of a hybrid bottomonium.

For an electric gluon charmonium hybrid, the calculated energy spectrum showed that the ground state appeared a few hundred MeV below the mass of Y(4260), while the energy of the first excited state was comparable to that of $\psi$(4415). Also, the calculated widths for the decays of the ground state or the first excited state of a $c\bar{c}g$ into $D^{(*)}\bar{D}^{(*)}$ were much larger than the total widths of these exotic meson candidates. For a magnetic gluon $c\bar{c}g$, the calculated ground state energy was comparable to the energy of the first excited state of an electric gluon $c\bar{c}g$, only confirming that the calculated ground state was consistent with the results of earlier studies. Our results supported the view that if one of these exotic meson candidates is a hybrid charmonium, its constituent gluon is magnetic.

In relation to a $c\bar{c}g$, similar conclusions could be reached regarding a bottomonium hybrid. The ground states of an electric gluon $b\bar{b}g$ appeared about 200 MeV below the mass of $\Upsilon(10860)$, while the first excited state was a few hundred MeV above the mass.

In addition, we evaluated the widths of the decays of the ground states and the first excited states of a $b\bar{b}g$ into $B^{(*)}\bar{B}^{(*)}$. We found that, for the ground states, some of the $B^{(*)}\bar{B}^{(*)}$ channels could not appear in our calculations and that the widths were large to the extent that it seems unlikely that $\Upsilon(10860)$ is the ground state of an electric gluon $b\bar{b}g$. Still, there could be a possibility that the first excited state is the hybrid meson candidate, given the fact that $\alpha_s$ used for the calculations of the decay widths could be smaller under more realistic conditions.

When it comes to the magnetic gluon, our calculations of the ground states of a $b\bar{b}g$ are above the lattice calculations but were in good agreement with the QCD sum rule analysis. This could be other indirect evidence to suggest that the gluonic excitations could appear in a bottomonium.

## ACKNOWLEDGMENTS

The authors thank Professor M. Oka for many useful comments and discussions. S. Y. is supported by the Grant-in-Aid for Scientific Research (Grants No. 25247036, No. 15K17641, and No. 17K05435) from Japan Society for the Promotion of Science and by the MEXT-Supported Program for the Strategic Foundation at Private Universities, "Topological Science" under Grant No. S1511006.

## APPENDIX A: THE COORDINATE SYSTEMS OF A THREE-BODY SYSTEM

This Appendix covers the hyperspherical and Jacobi coordinates for a three-body system which consists of particles 1, 2, and 3; their masses are denoted by $m_1$, $m_2$, and $m_3$, respectively.

For the present case, there exist three different types of Jacobi coordinates: X coordinates, Y coordinates, and T coordinates (see Fig. 9). In Fig. 9, $X_1$ is the distance between particles 1 and 3, and $Y_1$ is the distance between particle 2 and the center of mass of particles 1 and 3. Similar things hold true for $X_2$, $Y_2$ and $X_3$, $Y_3$. Picking up one of these coordinates, we can express the Jacobi coordinates by [37]

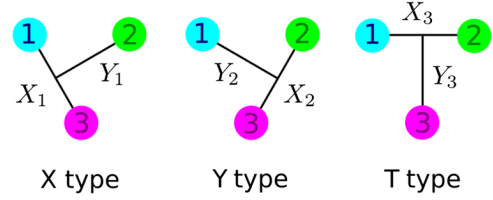

FIG. 9. The three different types of Jacobi coordinates.

$$\mathbf{x}_i = \sqrt{\frac{m_j m_k}{m_j + m_k}}(\mathbf{r}_j - \mathbf{r}_k), \quad (A1)$$

$$\mathbf{y}_i = \sqrt{\frac{m_i(m_j + m_k)}{M}}\left(\mathbf{r}_i - \frac{m_j \mathbf{r}_j + m_k \mathbf{r}_k}{m_j + m_k}\right), \quad (A2)$$

$$\mathbf{R} = \frac{m_1 \mathbf{r}_1 + m_2 \mathbf{r}_2 + m_3 \mathbf{r}_3}{M}, \quad (A3)$$

where $M = m_j + m_k + m_i$, and $(i, j, k)$ is a cyclic permutation of (1, 2, 3). Then, we note that the three different Jacobi coordinates (X, Y, and T) are connected to each other via the kinematic rotation,

$$\begin{pmatrix} \mathbf{x}_k \\ \mathbf{y}_k \end{pmatrix} = \begin{pmatrix} -\cos\phi_{ki} & \sin\phi_{ki} \\ -\sin\phi_{ki} & -\cos\phi_{ki} \end{pmatrix} \begin{pmatrix} \mathbf{x}_i \\ \mathbf{y}_i \end{pmatrix}, \quad (A4)$$

where $\phi_{ki}$ is the angle of the kinematic rotation. Specifically, the angle is $\phi_{ki} = \arctan(-1)^P \sqrt{\frac{M m_j}{m_i m_k}}$, where $P$ is determined according to the permutation of (1, 2, 3). $P$ is even/odd when $(k, i, j)$ is an even/odd permutation of (1, 2, 3).

When it comes to a hyperspherical-coordinate system, we introduce a quantity which does not depend on the three coordinates (X, Y, and T) by the following identities:

$$x_1^2 + y_1^2 = x_2^2 + y_2^2 = x_3^2 + y_3^2 = \rho^2, \quad (A5)$$

where the hyperradius $\rho$ has already been defined in (9). We note that from the definition of the kinematic rotation, it is apparent that $\{x_j^2 + y_j^2\}_{j=1,2,3}$ are invariant quantities under the rotation. Then, $x_i$ and $y_i$ are expressed as [38]

$$x_i = \rho \cos\alpha_i, \qquad y_i = \rho \sin\alpha_i, \quad (A6)$$

where $\{\alpha_i\}_{i=1,2,3}$ are hyperangles. By using the $\rho$ and $\alpha_i$, we can define the hyperspherical harmonics (11). Accordingly, it is possible to express the hyperspherical harmonics in three different ways: $\{\mathcal{Y}^{L_{x_i},L_{y_i}}_{KLM_L}(\Omega_{5,i})\}_{i=1,2,3}$. These different HHs are linked to each other via the Raynal-Revai coefficients: $\langle L_{x_k} L_{y_k} | L_{x_i} L_{y_i} \rangle_{KL}$ [38].

In addition to the coordinates $\mathbf{x}, \mathbf{y}$, and $\mathbf{R}$, we need to mention their conjugate momenta $\boldsymbol{\kappa}_x, \boldsymbol{\kappa}_y, \mathbf{P}$. They are given by [39]

$$\kappa_x = \sqrt{\frac{m_j m_k}{m_j + m_k}} \left( \frac{\mathbf{k}_j}{m_j} - \frac{\mathbf{k}_k}{m_k} \right), \quad (A7)$$

$$\kappa_y = \sqrt{\frac{m_i(m_j + m_k)}{M}} \left( \frac{\mathbf{k}_j + \mathbf{k}_k}{m_j + m_k} - \frac{\mathbf{k}_i}{m_i} \right)(-1), \quad (A8)$$

$$\mathbf{P} = \sqrt{\frac{1}{M}}(\mathbf{k}_1 + \mathbf{k}_2 + \mathbf{k}_3). \quad (A9)$$

The factor $(-1)$ for $\kappa_y$ does not exist in Jibuti's paper; this factor is just a matter of choice. Here, $\kappa$ is defined by $\kappa^2 = \kappa_x^2 + \kappa_y^2$ [39]. Relating to the hyperspherical coordinates, we have $\kappa_x = \kappa \cos \alpha_\kappa$ and $\kappa_y = \kappa \sin \alpha_\kappa$.

## APPENDIX B: FITTING PROCEDURES

In this Appendix, we show the basic fitting procedures with regard to the relevant parameters of a bottomonium. We determine the values of the string tension $\sigma$ and the color-Coulomb parameter $\alpha_s$ by fitting them to the energy spectrum of a $b\bar{b}$. First, we need to set the Hamiltonian $H_{\text{tot}}^{b\bar{b}}$ for the two-body system:

$$H_{\text{tot}}^{b\bar{b}} = H_0^{b\bar{b}} + V_{\text{Lin+Cou}}^{b\bar{b}} + V_{\text{str}}^{b\bar{b}} + V_{LS}^{b\bar{b}} + V_{\text{ss}}^{b\bar{b}} + V_{ST}^{b\bar{b}}, \quad (B1)$$

where $H_0^{b\bar{b}}$ is the free Hamiltonian for the bottomonium; $V_{\text{Lin+Cou}}^{b\bar{b}}$ is the term of the linear and color-Coulomb potentials; $V_{\text{str}}^{b\bar{b}}$ is a string correction term. In addition, we use the $LS$ coupling term $V_{LS}^{b\bar{b}}$, the spin-spin interaction $V_{\text{ss}}^{b\bar{b}}$, and the tensor term $V_{ST}^{b\bar{b}}$. These interactions are explicitly shown as follows:

$$V_{\text{str}}^{b\bar{b}} = -\frac{\sigma \mathbf{L}^2}{6\mu^2 r}(\hbar c),$$

$$V_{LS}^{b\bar{b}} = -\frac{\sigma}{2\mu^2 r}(\mathbf{L} \cdot \mathbf{S})(\hbar c) + \frac{2\alpha_s}{\mu^2 r^3}(\mathbf{L} \cdot \mathbf{S})(\hbar c)^3,$$

$$V_{\text{ss}}^{b\bar{b}} = \frac{32\pi \alpha_s}{9\mu^2}(\mathbf{s}_q \cdot \mathbf{s}_{\bar{q}})\delta(\mathbf{r})(\hbar c)^3,$$

$$V_{ST}^{b\bar{b}} = \frac{4\alpha_s}{3\mu^2 r^5}[3(\mathbf{s}_q \cdot \mathbf{r})(\mathbf{s}_{\bar{q}} \cdot \mathbf{r}) - r^2(\mathbf{s}_q \cdot \mathbf{s}_{\bar{q}})](\hbar c)^3,$$

where we explicitly write $\hbar c$ factors as well. We replace the delta function which appears in the spin-spin interaction with the following exponential function [40]:

$$\delta(\mathbf{r}) \to \frac{\Lambda^2}{4\pi r} e^{-\Lambda r}. \quad (B2)$$

We make this sort of modification in part because we cannot treat the delta function numerically; also, we need to address numerical instability issues relating to the delta function. In addition, we modify the tensor interaction as follows [40]:

TABLE VII. The calculated energy spectra and the measured one of a bottomonium. The parameters are $m_b = 5.00$ GeV and $\alpha_s = 0.55$ for set A, and $m_b = 4.85$ and $\alpha_s = 0.45$ for set B. $\Lambda = 6.0$ (1/fm). The experimental data are from [34].

|  |  | Set A (GeV) | Set B (GeV) | $E_{\text{exp}}$ (GeV) |
|---|---|---|---|---|
| $\chi_{b2}(1P)$ | $^3P_2$ | 10.176 | 9.964 | $9.91221 \pm (0.57 \times 10^{-3})$ |
| $h_b(1P)$ | $^1P_1$ | 10.147 | 9.945 | $9.8993 \pm (0.8 \times 10^{-3})$ |
| $\chi_{b1}(1P)$ | $^3P_1$ | 10.093 | 9.927 | $9.89278 \pm (0.57 \times 10^{-3})$ |
| $\chi_{b0}(1P)$ | $^3P_0$ | 10.014 | 9.893 | $9.85944 \pm (0.73 \times 10^{-3})$ |
| $\Upsilon(1S)$ | $^3S_1$ | 9.426 | 9.401 | $9.46030 \pm (2.6 \times 10^{-4})$ |
| $\eta_b(1S)$ | $^1S_0$ | 9.395 | 9.381 | $9.3990 \pm (2.3 \times 10^{-3})$ |

TABLE VIII. The relevant parameter sets for a bottomonium.

|  | $\sigma$ (GeV$^2$) | $\alpha_s$ | $\Lambda$ (fm$^{-1}$) | $m_b$ (GeV) | $\mu$ (GeV) |
|---|---|---|---|---|---|
| Set A | 0.16 | 0.55 | 6.0 | 5.00 | 5.42 |
| Set B | 0.16 | 0.45 | 6.0 | 4.85 | 5.16 |

$$V_{ST}^{b\bar{b}} \to \frac{\alpha_s(1 - e^{-\Lambda r})^2}{3\mu^2 r^3} S_{12}(\hbar c)^3, \quad (B3)$$

where $S_{12} = 12(\mathbf{s}_b \cdot \mathbf{n})(\mathbf{s}_{\bar{b}} \cdot \mathbf{n}) - 4(\mathbf{s}_b \cdot \mathbf{s}_{\bar{b}})$.

Second, we changed the mass $m_b$ of a bottom quark and calculated the energy spectrum, fixing $\sigma$ and $\alpha_s$. We used the auxiliary field method to determine the ground state and the value of a relevant auxiliary field. After obtaining these values, we calculated the higher states of a charmonium.

The results are shown in the set A column of Table VII, where we can see that the calculated energies agree with the experimental data for the bottom two states. When it comes to higher states, the difference between our values and the experimental data is more than 1.5%.

Then, we changed both $m_b$ and $\alpha_s$ and searched the parametrization under which the calculated energy well reproduces the $b\bar{b}$ energy spectrum. Accordingly we get $\alpha_s = 0.45$ and the corresponding energy spectrum which is shown in the set B column of Table VII. This parametrization produces the theoretical values that are consistent with the experimental results.

Finally, we present the summary of the parameter sets in Table VIII.

## APPENDIX C: LAGRANGE MESH TECHNIQUE

In this Appendix, we explain the Lagrange mesh method.
Suppose the Lagrange functions $\{f_j\}_{j=1,...,N}$ are real functions, and $\{z_j\}_{j=1,...,N}$ are their mesh points; the set of these points is the quadrature-point equivalent in the Gauss quadrature. Then, the Lagrange functions satisfy the following condition [41]:

$$f_i(z_j) = \frac{\delta_{ij}}{\sqrt{\lambda_i}}. \tag{C1}$$

We can easily check that the functions are orthonormal:

$$\langle f_i | f_j \rangle_G = \sum_{k=1}^{N} \lambda_k \frac{\delta_{ik}}{\sqrt{\lambda_i}} \frac{\delta_{jk}}{\sqrt{\lambda_j}} = \lambda_i \frac{\delta_{ji}}{\sqrt{\lambda_i \lambda_j}} \tag{C2}$$

$$= \delta_{ij}, \tag{C3}$$

where the brackets $\langle | \rangle_G$ stand for the Gauss quadrature calculation.

### 1. Scaling

In actual calculations, it is convenient to introduce scaling to reduce the computational cost. Here we denote the scaling parameter by $x_0$ such that $z = \frac{x}{x_0}$. Suppose the $T$ operator is $T = -\frac{d^2}{dx^2}$. Then, using the scaled parameter $z$ and the scaled Lagrange functions $\{f(x/x_0)\}_{j=1,\ldots,N}$, we can write it matrix element as

$$\int f_i^\star(x/x_0) T f_j(x/x_0) dx = x_0 \int f_i^\star(z) \left(-\frac{d}{d(x_0 z)^2}\right) f_j(z) dz$$

$$= x_0 T_{ij}, \tag{C4}$$

where the matrix element $T_{ij}$ contains the scaling parameter:

$$T_{ij} = \frac{1}{x_0^2} \int f_i^\star(z) \left(-\frac{d}{dz^2}\right) f_j(z) dz. \tag{C5}$$

We also calculate the matrix element of a potential $V(x)$ by this scaled functions:

$$\int f_i^\star(x/x_0) V(x) f_j(x/x_0) dx = x_0 V(x_0 z_i) \delta_{ij}. \tag{C6}$$

When it comes to the wave function, we expand it so as to keep the normalization

$$\psi(x) = \frac{1}{\sqrt{x_0}} \sum_{j=1}^{N} c_j f_j(x/x_0). \tag{C7}$$

It follows from these expressions that the equations of motion are written as

$$\sum_{j=1}^{N} (T_{ij} + V(x_0 z_i) \delta_{ij} - E \delta_{ij}) c_j = 0, \tag{C8}$$

where, for simplicity, the (reduced) mass parameter $\mu$ and the reduced Planck constant $\hbar$ are set to unity.

Also we note that in practice, the Gauss quadrature does not yield accurate matrix elements for Coulomb-type and centrifugal-type potentials [41]; in addition, we often use nonorthogonal Lagrange functions. For these reasons, we use the regularized functions which are given by

$$\tilde{f}_j(z) = \frac{z}{z_j} f_j(z) = (-1)^j \sqrt{\frac{1}{z_j h_n^{(\alpha)}}} z^{1+\frac{\alpha}{2}} e^{-\frac{z}{2}} \frac{L_n^{(\alpha)}(z)}{z - z_j}. \tag{C9}$$

This expression allows us to realize that the regularized functions still satisfy the condition (C1): $\tilde{f}_j(z_k) = \frac{z_k}{z_j} f_j(z_k)$. The shortcoming of the regularized functions is that they do not form an orthonormal set:

$$\langle \tilde{f}_i | \tilde{f}_j \rangle = \delta_{ij} + \frac{(-1)^{i-j}}{\sqrt{z_i z_j}}, \tag{C10}$$

where the brackets $\langle | \rangle$ mean the norm or matrix element calculation where no approximation is used. It could be noteworthy that we could treat the basis function like orthonormal functions because numerical experiments suggest that ignoring the nonorthogonality does not lead to a significant loss of accuracy [41].

The matrix elements of the double differential operator are expressed as

$$T_{ij} = \frac{1}{x_0^2} t_{ij}, \tag{C11}$$

$$t_{ij} = \begin{cases} \frac{(-1)^{i-j+1}}{4\sqrt{z_i z_j}} + (-1)^{i-j} \frac{z_i + z_j}{\sqrt{z_i z_j}(z_i - z_j)^2} & \text{for } i \neq j, \\ \frac{(-1)^{i-j+1}}{4\sqrt{z_i z_j}} - \frac{z_i^2 - 2(2N+\alpha+1)z_i + (\alpha^2 - 4)}{12 z_i^2} & \text{for } i = j. \end{cases} \tag{C12}$$

### 2. Salpeter-type Hamiltonian

We can use the Lagrange mesh technique to solve the eigenvalue problems, which involves the Salpeter Hamiltonian: $H = \sqrt{\mathbf{p}^2 + m^2} + V$. The procedure to calculate the matrix elements of $\sqrt{\mathbf{p}^2 + m^2}$ is as follows [42]:
(1) Calculate $P_{ij}^2 = \langle \tilde{f}_i | \mathbf{p}^2 + m^2 | \tilde{f}_j \rangle$
(2) Diagonalize $P_{ij}^2 = V D^2 V^{-1}$
(3) Derive $D$ by calculating the positive square root of each element of $D^2$
(4) Obtain the matrix elements through $P_{ij} = V D V^{-1}$

### 3. Solving inverse problems

After obtaining the wave function and the eigenenergy of a system, we use (C1) and (C7) to obtain the relation between the expansion coefficients and the wave function:

$$c_k = \sqrt{x_0 \lambda_k} \phi(x_k). \tag{C13}$$

This relation allows us to solve the inverse problem where the eigenenergy and the wave function are used as inputs. We substitute this relation into (C8) to have the expression for the potential [43] (we note that here the reduced Planck constant is set to unity):

$$V(x_i) = E - \sum_{j=1}^{N} T_{ij} \frac{\sqrt{\lambda_j}\phi(x_j)}{\sqrt{\lambda_i}\phi(x_i)}. \tag{C14}$$

## APPENDIX D: MOMENTUM REPRESENTATION

In this short Appendix, we briefly show the momentum representation of the wave function for a hybrid meson. This is used for the calculations of widths for the processes of $c\bar{c}g \to D^{(*)}\bar{D}^{(*)}$ and $b\bar{b}g \to B^{(*)}\bar{B}^{(*)}$. In relation to this, it is worth mentioning that the whole wave function of a quarkonium hybrid, which is originally expressed in the hyperspherical coordinates, is rewritten by using the Cartesian coordinates:

$$\Psi(\mathbf{X}, \mathbf{Y}) = \sum_{S_{\text{tot}}} R(X,Y)[[Y_{L_{q\bar{q}}} \otimes Y_{L_g}]_{L_{\text{tot}}} \otimes [|S_{q\bar{q}}\rangle \otimes |S_g\rangle]_{S_{\text{tot}}}]_{JM_J}, \tag{D1}$$

where the relating subscripts are omitted in $R(X,Y)$ for simplicity.

After that, we use the Rayleigh expansion to conduct Fourier transformation:

$$e^{i(\mathbf{K}_1\cdot\mathbf{r}_1+\mathbf{K}_2\cdot\mathbf{r}_2)} = (2\pi)^3 \sum_{L_1,L_2,m_1,m_2} i^{L_1}i^{L_2}\frac{1}{\sqrt{K_1 r_1}}\cdot\frac{1}{\sqrt{K_2 r_2}} J_{L_1+1/2}(K_1 r_1) J_{L_2+1/2}(K_2 r_2) Y^*_{L_1 m_1}(\hat{\mathbf{K}}_1) Y_{L_1 m_1}(\mathbf{r}_1) Y^*_{L_2 m_2}(\hat{\mathbf{K}}_2) Y_{L_2 m_2}(\mathbf{r}_2). \tag{D2}$$

Then, we obtain the expression for the momentum representation $\Phi_{JM_J}$ of the whole wave function:

$$\Phi_{JM_J}(\mathbf{p}_{q\bar{q}}, \mathbf{k}_g) = \sum_{S_{\text{tot}}} (2\pi)^3 \tilde{R}(p_{q\bar{q}}, k_g) [[Y_{L_{q\bar{q}}} \otimes Y_{L_g}]_{L_{\text{tot}}} \otimes [|S_{q\bar{q}}\rangle \otimes |S_g\rangle]_{S_{\text{tot}}}]_{JM_J} \tag{D3}$$

$$= \sum_{S_{\text{tot}}} (2\pi)^3 \tilde{R}(p_{q\bar{q}}, k_g) \sum_{J_{q\bar{q}}, J_g} X(L_{q\bar{q}} L_g L_{\text{tot}}, S_{q\bar{q}} S_g S_{\text{tot}}, J_{q\bar{q}} J_g J) [[Y_{L_{q\bar{q}}} \otimes |S_{q\bar{q}}\rangle]_{J_{q\bar{q}}} \otimes [Y_{L_g} \otimes |S_g\rangle]_{J_g}]_{JM_J}, \tag{D4}$$

where $X(abc, def, ghi)$ is defined by

$$X(abc, def, ghi) = \sqrt{(2c+1)(2f+1)(2g+1)(2h+1)} \begin{Bmatrix} a & b & c \\ d & e & f \\ g & h & i \end{Bmatrix}. \tag{D5}$$

Finally, putting

$$\hat{\Phi}_{J_{q\bar{q}}J_g}(\mathbf{p}_{q\bar{q}}, \mathbf{k}_g) = \sum_{S_{\text{tot}}} (2\pi)^3 \tilde{R}(p_{q\bar{q}}, k_g) X(L_{q\bar{q}} L_g L_{\text{tot}}, S_{q\bar{q}} S_g S_{\text{tot}}, J_{q\bar{q}} J_g J) [[Y_{L_{q\bar{q}}} \otimes |S_{q\bar{q}}\rangle]_{J_{q\bar{q}}} \otimes [Y_{L_g} \otimes |S_g\rangle]_{J_g}]_{JM_J}, \tag{D6}$$

we express $\Phi_{JM_J}(\mathbf{p}_{q\bar{q}}, \mathbf{k}_g)$ as

$$\Phi_{JM_J}(\mathbf{p}_{q\bar{q}}, \mathbf{k}_g) = \sum_{J_{q\bar{q}}J_g} \hat{\Phi}_{J_{q\bar{q}}J_g}(\mathbf{p}_{q\bar{q}}, \mathbf{k}_g). \tag{D7}$$